\documentstyle[12pt]{article}
\def\newline{\relax\ifhmode\null\hfil\break\else\nonhmodeerr@\newline\fi}
\def\frac#1#2{{#1\over#2}}
\def\text#1{{\hbox{\rm #1}}}
\def\flushpar{{\par \noindent}}

\def\csb{\chi SB}
\newcommand{\beq}{\begin{equation}}
\newcommand{\eeq}{\end{equation}}
\newcommand{\bea}{\begin{eqnarray}}
\newcommand{\eea}{\end{eqnarray}}
\def\Id{ \mbox{1\hspace{-1.2mm}I} }
\def\BE{\begin{equation}}
\def\EE{\end{equation}}
\def\BA{\begin{eqnarray}}
\def\EA{\end{eqnarray}}
\def\BAN{\begin{eqnarray*}}
\def\EAN{\end{eqnarray*}}

\def\nn{\nonumber\\}

\def\tr{\mbox{tr}}

\def\det{\mbox{det}}

\def\gm5{\gamma^5}

\begin{document}
\thispagestyle{empty}
\begin{flushright}
NTUTH-98-101 \\
October 1998
\end{flushright}
\bigskip\bigskip\bigskip
\vskip 2.5truecm
\begin{center}
{\LARGE {GW fermion propagators and chiral condensate}}
\end{center}
\vskip 1.0truecm
\centerline{Ting-Wai Chiu}
\vskip5mm
\centerline{Department of Physics, National Taiwan University}
\centerline{Taipei, Taiwan 106, R.O.C.}
\centerline{\it E-mail : twchiu@phys.ntu.edu.tw}
\vskip 2cm
\bigskip \nopagebreak \begin{abstract}
\noindent

By exploiting the chiral symmetry, we derive analytic formulas for
the massless and the massive Ginsparg-Wilson fermion propagators.
Using these formulas, we derive an expression for the chiral condensate
which is the order parameter for spontaneous chiral symmetry
breaking in QCD. These formulas provide the proper way to compute the
fermion propagators and may save a significant amount of computing time
especially for large lattices in four dimensions.


\end{abstract}
\vskip 1.5cm

\newpage\setcounter{page}1

It is well understood that the proper way to break the chiral symmetry
of a chirally symmetric Dirac operator $ D_c $ on the lattice is to replace
$ D_c $ by
\beq
D = D_c ( \Id + R D_c )^{-1}
\label{eq:gwf}
\eeq
where $ R $ is any hermitian operator which is local in the position space
and trivial in the Dirac space. Then the Dirac operator $ D $ satisfies
the Ginsparg-Wilson relation \cite{gwr}
\beq
\label{eq:gwo}
D \gamma_5 + \gamma_5 D = 2 D \gamma_5 R D
\eeq
where the particular form of the chiral symmetry breaking on the RHS
having one $ \gamma_5 $ sandwiched by two Dirac operators is the signature
of the Ginsparg-Wilson fermion. This implies the chiral Ward identities,
non-renormalization of vector and flavor non-singlet axial vector
currents, and non-mixing of operators in different chiral representations,
as shown by Hasenfratz \cite{ph98:2}. The general solution
\cite{twc98:6a,twc98:9a} of the GW relation, Eq. (\ref{eq:gwf}), can be
regarded as a chiral symmetry breaking transformation which gaurantees that
a zero mode of $ D_c $ is also a zero mode of $ D $ and vice versa, hence
the index of $ D $ is equal to the index of $ D_c $. Then it is obvious
to see that the presence of $ R $ cannot produce a zero mode for $ D $
if $ D_c $ does not possess any zero modes in topologically nontrivial
background gauge fields. We refer to ref. \cite{twc98:9a,twc98:10a} for
further discussions on topological characteristics of $ D $ ( $ D_c $ ).
Now we can replace $ R $ in Eq. (\ref{eq:gwo}) by any two
hermitian operators $ S $ and $ T $ having the same properties of $ R $,
and rewrite Eq. (\ref{eq:gwo}) as the following,
\bea
\label{eq:chi_sym_lat}
D \gamma_5 ( \Id - S D ) + ( \Id - D T ) \gamma_5 D = 0 \ ,
\eea
then it is evident that the action $ A = \bar\psi D \psi $ is invariant
under the finite chiral transformation on the lattice
\bea
\label{eq:ct1}
\bar\psi \rightarrow \bar\psi \exp [ i \theta (\Id- D T) \gamma_5 ] \\
\label{eq:ct2}
\psi \rightarrow \exp [ i \theta  \gamma_5 ( \Id - S D ) ] \psi,
\eea
where $ \theta $ is a global parameter. This generalizes the
infinitesimal chiral transformation observed by L\"uscher \cite{ml98:2}.
In general, we can regard the Ginsparg-Wilson relation in the form of
Eq. (\ref{eq:chi_sym_lat}) as the exact chiral symmetry on the lattice,
which recovers the usual chiral symmetry $ \gamma_5 D + D \gamma_5 = 0 $
in the continuum limit ( $ a \to 0 $ ). Such a generalization seems to be
cruical for formulating chiral gauge theories on the lattice.

In this paper, we will concentrate on the problem of evaluating the
fermion propagator $ D^{-1} $ of the Ginsparg-Wilson fermion.
From Eq. (\ref{eq:gwf}), we immediately obtain
\bea
\label{eq:Di}
D^{-1} = D_c^{-1} + R
\eea
Hence, the fermion propagator $ D^{-1} $ is completely determined for
any $ R $ if the chiral fermion propagator $ D_c^{-1} $ has been
evaluated. In fact, the long distance properties of $ D^{-1} $ are
governed by the long distance properties of $ D_c^{-1} $ since $ R $ is
supposed to be local. Furthermore, we will show that the short
distance behavior of $ D^{-1} $ which is relevant to the
the chiral condensate is also determined by $ D_c^{-1} $.
Therefore the physics of any Ginsparg-Wilson fermion
operator $ D $ actually lies in its chiral limit $ D_c $. However,
only $ D_c $ itself is not sufficent to constitute the correct lattice
fermion operator. According to the Nielson-Ninomiya no-go theorem \cite{no-go},
if we require $ D $ to be local and free of species doubling, then the chiral
symmetry $ \{ \gamma_5, D \} = 0 $ must be broken. The transformation
in Eq. (\ref{eq:gwf}) provides the proper chiral symmetry breaking
which not only preserves the correct physics of $ D_c $,
but also removes its non-locality for any gauge configurations
as well as its singularities in topologically nontrivial gauge fields.
We note in passing that the locality requirement for the GW fermion should
be imposed on the unitary operator $ V $ in Eq. (\ref{eq:VDc}) rather than
on $ D $, since the non-locality of $ V $ is due to the initial construction
and thus cannot be diminished, while the locality of $ D $ is always
gauranteed by a properly chosen $ R $ in Eq. (\ref{eq:gwf}).

In the following, we derive a formula for the chiral fermion propagator
for any $ D_c $ satisfying the hermiticity condition
\bea
\gamma_5 D_c \gamma_5 = D_c^{\dagger} \ .
\label{eq:hermit}
\eea
The hermiticity condition and the chiral symmetry of $ D_c $ implies that
$ D_c $ is antihermitian, thus there exists one to one correspondence
between $ D_c $ and a unitary operator $ V $,
\bea
D_c = (\Id + V )(\Id - V )^{-1}, \hspace{4mm}
V = (D_c - \Id)( D_c + \Id)^{-1}.
\label{eq:VDc}
\eea
where $ V $ also satisfies the hermiticity condition
( $ \gamma_5 V \gamma_5 = V^{\dagger} $ ).
Then the unitary operator $ V $ can be expressed in terms of
a hermitian operator $ h $,
\bea
\label{eq:V5h}
V = \gamma_5 h = \left( \begin{array}{cc}
                         \Id  &    0    \\
                           0  & -\Id
                        \end{array}      \right)
                  \left( \begin{array}{cc}
                          h_1            &  h_2    \\
                          h_2^{\dagger}  &  h_3
                        \end{array}      \right)
               =  \left( \begin{array}{cc}
                          h_1            &  h_2    \\
                         -h_2^{\dagger}  & -h_3
                        \end{array}      \right)
\eea
where $ h_1^{\dagger} = h_1 $ and $ h_3^{\dagger} = h_3 $.
Using the unitarity condition $ V^{\dagger} V = \Id $,
we have $ h^2 = \Id $,
\bea
\label{eq:h2}
h^2 = \left( \begin{array}{cc}
     h_1^2 + h_2 h_2^{\dagger}              &   h_1 h_2 + h_2 h_3    \\
     h_2^{\dagger} h_1 + h_3 h_2^{\dagger}  &   h_2^{\dagger} h_2 + h_3^2
            \end{array}      \right)
    = \left( \begin{array}{cc}
              \Id          &  0    \\
                0          &  \Id
             \end{array}      \right)
\eea
Then we immediately obtain
\bea
D_c
\label{eq:Dc}
= (\Id + V )(\Id - V )^{-1}
=  \left[ \begin{array}{cc}
                     0                 &  (\Id - h_1 )^{-1} h_2  \\
     -h_2^{\dagger} (\Id - h_1)^{-1}   &   0
              \end{array}                           \right]
\eea
and
\bea
\label{eq:Dci}
D_c^{-1}
= (\Id - V )(\Id + V )^{-1}
=  \left[ \begin{array}{cc}
                      0                 &  -(\Id + h_1 )^{-1} h_2  \\
       h_2^{\dagger} (\Id + h_1)^{-1}   &   0
                   \end{array}                           \right]
\eea
The last formula together with Eq. (\ref{eq:Di}) reduce the task of
computing $ D^{-1} $ for {\it any} $ R $ to that of computing
$ (\Id + h_1 )^{-1} $ only. For a four dimensional lattice of volume
$ \Omega = L^4 $, the former is to compute the inverse of a complex matrix of
size $ 4 L^4 \times 4 L^4 $ ( multiplicities due to the color and the flavor
have been suppressed ), while the latter only amounts to computing the inverse
of a hermitian matrix of size $ 2 L^4 \times 2 L^4 $.
Further simplifications are possible,
however, depend on the actual form of $ V $ ( $ D_c $ ) as well as the
algorithm used for computing the inverse. We do not intend to go into
these technical details in this paper. In principle, the formula
(\ref{eq:Dci}) provides the proper way to compute the fermion propagator for
any Ginsparg-Wilson fermion and serves as a starting point for devising
highly optimized algorithms for computing the fermion propagators.

In order to investigate the spontaneous chiral symmetry breaking
( $ \csb $ ) of QCD, we usually introduce a mass parameter $ m $ to the
Dirac operator $ D $, and then measure the chiral condensate in the
thermodynamic limit $ \Omega \to \infty $ followed by the massless limit
$ m \to 0 $. The proper way to introduce the mass parameter to the Dirac
operator $ D $ is to add the mass $ m $ to the chirally symmetric $ D_c $
as in the continuum theory, but {\it not } to $ D $ directly.
Then the GW-Dirac operator (\ref{eq:gwf}) becomes
\bea
D = ( D_c + m ) ( \Id + R D_c )^{-1}
\label{eq:gwfm}
\eea
It should be noted that the mass $ m $ is not added to the $ D_c $
in the last inverse operator which is inserted for breaking the
chiral symmetry {\it a la} Ginsparg-Wilson.
Then the fermion propagator for massive GW fermion is
\bea
\label{eq:DIM}
D^{-1} = R + (\Id - m R )( D_c + m )^{-1}
\eea
where
\bea
\label{eq:Dcmi}
( D_c + m )^{-1}
=  \left( \begin{array}{cc}
   m ( \Id - h_1 ) B  &  -B h_2  \\
     h_2^{\dagger} B  &  m^{-1}[\Id - h_2^{\dagger} B (\Id - h_1)^{-1} h_2 ]
                   \end{array}                           \right)
\eea
and
\beq
\label{eq:B}
B = [ \Id + m^2 + ( \Id - m^2 ) h_1 ]^{-1} \ .
\eeq
Again, the amount of computations using above formulas is significantly
less than that of computing $ (D_c + m )^{-1} $ directly.
Once $ ( D_c + m )^{-1} $ is evaluated, then it almost takes no time
to obtain $ D^{-1} $ for any $ R $ using Eq. (\ref{eq:DIM}).

Now we can write down the usual expression for the chiral condensate
\bea
\label{eq:chi}
\chi = \lim_{m \to 0 } \lim_{ \Omega \to \infty } \frac{1}{\Omega}
\left< \sum_{x} \mbox{tr} \left[ \left. D^{-1}(x,x) \right |_{R=0} \right]
\right>
\eea
where the brackets $ < \cdots > $ denote
averaging over each gauge configuration with the weight \
$ \mbox{det}(D) \exp(-A_g) $, and $ A_g $ is the pure gauge action.
The reason to set $ R = 0 $ in $ D^{-1} $ is obvious,
since it is an arbitrary operator introduced for breaking
the chiral symmetry {\it a la} Ginsparg-Wilson, thus it is irrelevant
to the physical $\csb$ of QCD, and must be dropped
\footnote{Note that setting $ R=0 $ in (\ref{eq:chi}) does not necessarily
  reinstall doubling if $ D_c $ is non-local, as have been discussed in
  detail in ref. \cite{twc98:9a}.}.
Inserting (\ref{eq:DIM}) into (\ref{eq:chi}), we have
\bea
\label{eq:chi1}
\chi = \lim_{m \to 0 } \lim_{ \Omega \to \infty } \frac{1}{\Omega}
\left< \sum_{x} \mbox{tr}[  (D_c + m )^{-1}(x,x) ] \right>
\eea
The subtleness of Eq. (\ref{eq:chi1}) lies in its two limiting processes.
If the order of these two limiting processes is exchanged, then $ \chi $
must be zero since spontaneous chiral symmetry breaking could not occur
on a finite lattice for exactly massless fermion.
Using (\ref{eq:Dcmi}), we can simplify
Eq. (\ref{eq:chi1}) to the following expression
\bea
\label{eq:chi2}
\chi
= \lim_{m \to 0 } \lim_{ \Omega \to \infty } \frac{1}{\Omega}
  \left< \sum_{x} \ 2 m \ \tr \left[ (\Id - h_1)B \right](x,x) \right>
\eea
where $ B $ is defined in Eq. (\ref{eq:B}).
This is a simpler expression than (\ref{eq:chi1}) since
it only involves $ h_1 $, the hermitian submatrix
of $ h = \gamma_5 V $ defined in (\ref{eq:V5h}). The most time
consuming operation in (\ref{eq:chi2}) is to evaluate
the inverse matrix $ B $.

Now we restore the weight \ $ \mbox{det}(D) \exp(-A_g) $ \ embedded
inside the brackets $ < \cdots > $ and rewrite (\ref{eq:chi2}) in the
following,
\bea
\label{eq:chi3}
\chi
= \lim_{m \to 0 } \lim_{ \Omega \to \infty } \frac{1}{\Omega}
  \frac{1}{Z} \int [dU] \exp(-A_g) \ \det(D)
  \sum_{x} \ 2 m \ \mbox{tr} [ (\Id - h_1) B ] (x,x)
\eea
where
\beq
\label{eq:Z}
Z = \int [dU] \exp(-A_g) \ \det(D)
\eeq
and
\beq
\label{eq:detD}
\det(D) = \det(D_c) \ \det( \Id + R D_c )^{-1}
\eeq
using Eq. (\ref{eq:gwf}).
From the GW relation (\ref{eq:gwo}), it is obvious that we can
redefine $ D'= R D $ such that $ D' $ satisfies (\ref{eq:gwo})
with $ R = 1 $. However, one cannot set $ R=1 $ to the valence quarks
since that would produce terms which are not relevent to the physical
chiral symmetry breaking in QCD, as discussed after Eq. (\ref{eq:chi}).
On the other hand, for quarks in the internal quark loops,
the chiral symmetry must be broken such that $ D $ can be local
and free of species doubling, then the fermionic determinant and
the axial anomaly can turn out to be correct.
Therefore we can set $R$ to one in Eq. (\ref{eq:detD}),
provided that $ D_c $ has the correct classical continuum limit.
Using Eq. (\ref{eq:Dc}), we obtain
\beq
\label{eq:det}
\det(D) = \det \left( \frac{\Id + h_1 }{2} \right)
        = \det \left( \frac{\Id - h_3 }{2} \right)
\eeq
In topologically nontrivial gauge sectors, the zero modes of $ D $
corresponding to the $ -1 $ eigenmodes of $ h_1 $, but the $ +1 $
eigenmodes of $ h_3 $. Further implementations of computing $ \chi $
depend on the actual form of $ V $ ( or $ D_c $ ) as well as the simulation
algorithms. However, we will not go into these technical details in this
paper. Now we put the results together and rewrite the chiral
condensate in the following formula
\bea
\label{eq:xsb}
\chi
&=& \lim_{m \to 0 } \lim_{ \Omega \to \infty } \frac{1}{\Omega} \frac{1}{Z}
\int [dU] \ \exp(-A_g) \ \det \left( \frac{\Id + h_1 }{2} \right) \nn
& & \sum_{x} \ 2m \
\mbox{tr} \left[ (\Id - h_1)[\Id+m^2+(\Id-m^2)h_1]^{-1} (x,x) \right]
\eea
where $ h_1 $ is the hermitian submatrix defined in Eq. (\ref{eq:V5h}) and
\beq
Z = \int [dU] \exp(-A_g) \ \det \left( \frac{\Id + h_1 }{2} \right)
\eeq
If the chirally symmetric $ D_c $ is initially specified as
\bea
\label{eq:DcLR}
D_c
= \left( \begin{array}{cc}
             0  &  -D_L^{\dagger}  \\
           D_L  &  0
          \end{array}                           \right)
\eea
but not in terms of $ V = \gamma_5 h $, then it is straightforward to
rewrite all formulas for the fermion propagators and the chiral
condensate in terms of $ D_L $.

Finally, we note that Neuberger \cite{hn98:7} recently obtained an expression
for the chiral condensate in the massless limit and in a fixed gauge
background, in the framework of the overlap formalism \cite{rn95}.
Neuberger's result would agree with our general formula (\ref{eq:xsb}) if
the mass parameter is introduced to the fermion propagator and the formula
(\ref{eq:Dcmi}) is used.

In summary, we have derived analytic formulas for the GW fermion
propagators, for the massless and the massive cases respectively.
These formulas provide the proper way to compute GW fermion propagators
in analytic studies as well as numerical ones. They also serve as a
starting point for devising optimized algorithms for
computing fermion propagators. An analytic formula for the chiral
condensate is derived and is ready for numerical implementations for
Monte Carlo simulations including dynamical fermions.

\bigskip
\bigskip

\flushpar
{\bf Acknowledgement }
\bigskip

\noindent
This work was supported by the National Science Council, R.O.C.
under the grant number NSC88-2112-M002-016.

\vfill\eject

\vfill\eject

\end{document}